\documentclass[dvips]{article}
\usepackage{icrctc07}

\def\hess{H.E.S.S.\ }
\def\hessns{H.E.S.S.}
\def\mbh{M_{\mathrm{BH}}}
\def\mbhrm{M_{\mathrm{BH}}}
\def\gr{$\gamma$-ray\ }
\def\grs{$\gamma$-rays\ }

\title{A search for Very High Energy \gr emission from Passive Super-massive Black Holes}
\shorttitle{A search for VHE emission from passive SMBH}

\authors{G.Pedaletti$^{1,2}$, S. Wagner$^1$, W. Benbow$^3$, for the \hess collaboration}
\shortauthors{Pedaletti et al.}
\afiliations{$^1$Landessternwarte, Universit\"{a}t Heidelberg, K\"{o}nigstuhl, 69117 Heidelberg, Germany\\ 
$^2$ Max-Planck-Institut f\"{u}r Astronomie, K\"{o}nigstuhl, 69117 Heidelberg, Germany \\$^3$Max-Planck-Institut f\"{u}r Kernphysik, PO Box 103980, 69029 Heidelberg, Germany}
\email{gpedalet@lsw.uni-heidelberg.de}

\abstract{
Jets of Active Galactic Nuclei (AGN) are established emitters of very high energy (VHE; $>$100 GeV) $\gamma$-rays. VHE radiation is also expected to be emitted from the vicinity of super-massive black holes (SMBH), irrespective of their activity state. Accreting SMBH rotate and generate a dipolar magnetic field. In the magnetosphere of the spinning black hole, acceleration of particles can take place in the field gaps. VHE emission from these particles is feasible via leptonic or hadronic processes. Therefore quiescent systems, where the lack of a strong photon field allows the VHE emission to escape, are candidates for emission.  The \hess experiment has observed the passive SMBH in the nearby galaxy NGC 1399. No VHE \gr signal is observed from the galactic nucleus. Constraints set by the NGC 1399 observations are discussed in the context of different mechanisms for the production of VHE \gr emission.}

\begin{document}

\maketitle


\section{Introduction}

Spheroidal systems (such as elliptical galaxies, lenticular galaxies and early-type spiral galaxies with bulges) are commonly believed to host in the central region super-massive black holes with masses in the range $\mbh= 10^6 - 10^9 M_\odot$ \cite{rich}. During the early stages of galaxy evolution these SMBH accrete matter at high rates and are observed as bright QSOs. The radiative output at low energy (e.g. optical) decays from redshift z$>$3 to z=0 by almost 2 orders of magnitude. Therefore, the majority of SMBH in the local universe are not embedded in dense radiation fields. This enables VHE \grs to escape from the nuclear region without suffering from strong absorption via $\gamma$-photon pair absorption.
Several models \cite{gal-cen,lev,neron,slane} are proposed for the production of VHE \grs emission from these passive AGN. In all cases a large mass of the central object is the most important characteristic for generating a high VHE flux. \hess has already observed nine nearby galaxies whose black hole mass is measured \cite{mag,pel}. Only the case of NGC 1399 is considered here. Constraints on the physical parameters of the system (e.g. the magnetic field \textbf{B}) are derived using several of the aforementioned models.
\section{Acceleration Mechanism}
If the central black hole is accreting matter from a disk that also carries magnetic flux, it will develop a magnetosphere similar to those surrounding neutron stars. If the charge density is not too high in the magnetosphere of the spinning black hole, it is possible to have a non-zero component of the electric field \textbf{E} parallel to the magnetic field \textbf{B}. In this configuration field gaps are created, where acceleration of particles can take place \cite{slane}.

Various methods can be used to estimate the magnetic field B. For example, B is estimated:
\begin{itemize}
 \item assuming equipartition 
\begin{equation}\label{eqn:magn}
\frac{B^2}{8\pi} = \frac{1}{2} \rho(r_0)v^2_r(r_0),
\end{equation}
where $\rho$ is the mass density and $v_r$ is the radial infall velocity of the accreting matter (both being a function of $r_0$, the distance to the inner edge of the disk);
 \item from the angular momentum as in \cite{bick}
$$ B = 3.1\times10^3 \frac{\dot{m}^{1/2}}{M_{10}^{1/2}}\left(\frac{r}{r_\mathrm{g}}\right)^{-5/4} \textrm{Gauss,}$$
where $\dot{m}$ is the mass accretion rate in units of the Eddington mass accretion rate, $r_\mathrm{g}$ is the gravitational radius of the black hole and $M_{10}=(\mbhrm/10^{10}M_{\odot})$.
\end{itemize}

In the model of \cite{slane} protons accelerated in the outer part of the black hole magnetosphere will collide with other protons present in the accretion disk producing pions some of which decay into VHE \grs.
The available power is
\begin{equation}\label{eq:power}
 W_{\mathrm{max}} \sim 10^{27} \left(\mbh\right)^2 \left(B_4\right)^2 \textrm{ ergs s}^{-1}, 
\end{equation}

where $B_4=(B/10^4 \textrm{ Gauss})$.
Here it is assumed that the magnetic energy density is in equipartition with the accretion energy density, which depends on various properties of the accretion disk (see Eq.~\ref{eqn:magn}). 

In other models \cite{gal-cen,lev,neron} VHE \grs originate from electromagnetic processes such as synchrotron or curvature emission. Following the analogous arguments given in \cite{gal-cen} for the Galactic Center, synchrotron emission is not feasible due to a cut-off for protons and electrons at $\epsilon_{\gamma,\mathrm{max}} \simeq 0.3$ TeV and $\epsilon_{\gamma,\mathrm{max}} \simeq 0.16$ GeV respectively. These cut-offs are independent of the magnetic field strength. 
The energy of curvature photons (when curvature losses are the dominant ones) does not depend on the mass of the particle, so it is the same for electron or proton originated photons. The emission spectrum from curvature radiation can extend up to VHE energies, with a cut-off at:
\begin{equation}\label{eq:cutoff}
 E_{\mathrm{max}}\simeq14\left(M_{10}\right)^{1/2}\left(B_4\right)^{3/4} \textrm{TeV.}
\end{equation}

\section{VHE Observation of NGC 1399}

The giant elliptical galaxy NGC 1399 is located in the central region of the Fornax cluster at a distance of $20.3$ Mpc. An SMBH of $ \mbh = 1.06\times10^9 M_\odot$ resides in the central region. 
The nucleus of this galaxy is well known for its low emissivity at all wavelengths \cite{oconn}. Considering also the visibility of candidate sources for \hessns, NGC 1399 therefore emerged as the best candidate for this study.

\begin{figure}[htbp]
\centering
  \includegraphics[width=0.5\textwidth]{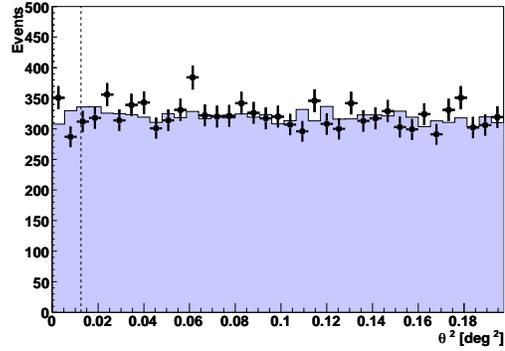}
  \caption{\small Distribution of squared angular distance from NGC 1399 for gamma-ray-like events in the ON region (dots) and in the OFF region (filled area, normalized). The dotted line represents the cut for point-like sources. Preliminary.  \normalsize}
 \label{fig:thetasq}
\end{figure}

\begin{figure}[htbp]
\centering
  \includegraphics[width=0.5\textwidth]{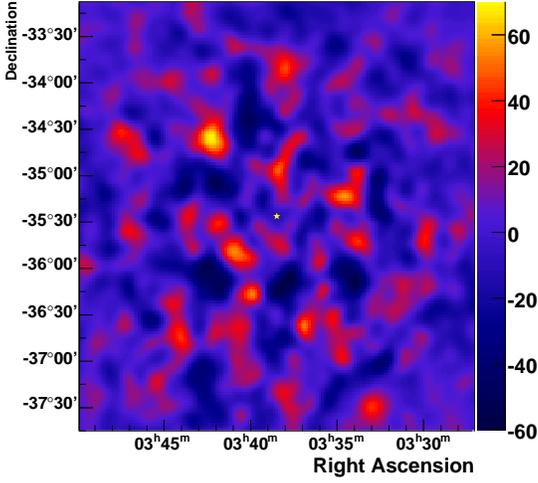}
  \caption{\small The smoothed (smoothing radius r=0.09˚) VHE excess in the region centered on NGC 1399. The yellow star indicates the position of the optical centre of NGC 1399. Preliminary.\normalsize}
 \label{fig:skymap}
\end{figure}

NGC 1399 was observed with the \hess array of imaging atmospheric-Cherenkov telescopes for a total of 22.4 h (53 runs of $\sim$28 min each). After applying the standard \hess data-quality selection criteria a total of 13.9 hours live time remain. The mean zenith angle is $Z_{\mathrm{mean}} = 22^\circ$. 
The data were reduced using the standard analysis tools and selection cuts \cite{benb} and the Reflected-Region method \cite{berge} for the estimation of the background. This leads to a post-analysis threshold of 200 GeV at $Z_{\mathrm{mean}}$. No significant excess (-29 events, -1$\sigma$) is detected from NGC 1399 (see Fig. \ref{fig:thetasq} and Fig.~\ref{fig:skymap}). Results are consistent with independent analysis in the collaboration.

Assuming a photon index of $\Gamma$=2.6, the upper limit (99$\%$ confidence level; \cite{fc}) on the integral flux above 200 GeV is:
$$I \left(>200\textrm{GeV}\right) < 2.3 \times 10^{-12} \textrm{ cm}^{-2}\textrm{s}^{-1},$$
or ~1\% of the Crab Nebula flux.
\section{Constraints from  NGC 1399 Observations}

\begin{figure}[htbp]
\centering
  \includegraphics[width=0.5\textwidth]{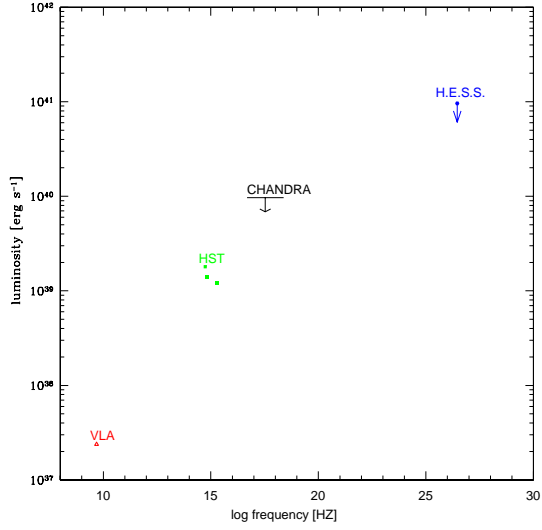}
  \caption{\small The SED of NGC 1399. All the data are for the core region. The archival points are VLA radio data (red triangles; \cite{sadl}), HST optical data (green squares; \cite{oconn}), and Chandra X-ray upper limits (solid line; \cite{lowen}). The blue dot is the  H.E.S.S. upper limit derived from the 2005 observations. Preliminary. \normalsize}
 \label{fig:SED}
\end{figure}

As can be seen from the spectral energy distribution (SED) of NGC 1399 in Fig.~\ref{fig:SED}, the VHE fraction of its total energy budget is potentially not-negligible. The \hess limit on the isotropic VHE \gr luminosity is:
$$L_\gamma < 9.6 \times 10^{40} \textrm{ erg s}^{-1}.$$
Here it is assumed that the \gr emission originates solely from the nucleus, even though the entire galaxy is point-like considering the angular resolution of \hess

In the case of NGC 1399 photon-photon pair absorption would not hide any possible VHE emission. The cross section $\sigma_{\gamma\gamma}$  of this process depends on the product of the energies of the colliding photons. In the case of VHE photons, the most effective interaction is with background photons of energy:
$$\epsilon_{\mathrm{IR}} \approx \left(E/\mathrm{1TeV}\right)^{-1} \textrm{ eV.} $$

The optical depth resulting from this absorption, in a source of luminosity $L$ and radius $R$, reads:

\begin{eqnarray}
\tau\left(E,R_{\mathrm{IR}}\right) & = & \frac{L_{\mathrm{IR}}\sigma_{\gamma\gamma}}{4\pi R_{\mathrm{IR}} \epsilon_{\mathrm{IR}}}
\nonumber\\
& \simeq & 1 \left[\frac{L_{\mathrm{IR}}\left(\epsilon \right)}{10^{-7} L_{\mathrm{Edd}}}\right] \left[\frac{R_\mathrm{S}}{R_{\mathrm{IR}}} \right]\left[\frac{E}{\textrm{1 TeV}}\right],
\nonumber
\end{eqnarray}
where $R_\mathrm{S}$ is the Schwarzschild radius of the black hole and $L_{\mathrm{Edd}}$ is the Eddington luminosity.
In the system here presented, the visibility of a 200 GeV photon requires  $L_{\mathrm{IR}} <7.9 \times 10^{40} \textrm{ ergs s}^{-1}$, a condition that seems to be satisfied.

In the p-p interaction scenario, assuming that all the available power (Eq. \ref{eq:power}) will be radiated in the VHE domain, the following limit for the magnetic field is obtained from the H.E.S.S. result:
$$ B < 92.6 \textrm{ Gauss.}$$

In order to maintain gaps in the magnetosphere, as is essential for particle acceleration, pair production should be avoided. Translating this condition into an upper limit for the magnetic field yields: 
$$B < 3.6 \times 10^4 \left(M_{10}\right)^{-2/7} = 6.8\times10^4 \textrm{ Gauss.}$$

Therefore the H.E.S.S. NGC 1399 data allow plausible values of the magnetic field.
Considering the production of a 1 TeV photon via curvature emission (Eq. \ref{eq:cutoff}) requires in the case of NGC 1399:
$$B=1.3\times10^3 \textrm{ Gauss.} $$
The non-detection of NGC 1399 does not constrain the magnetic field. In all the aforementioned scenarios, hadronic and/or leptonic, no clear constraints on the magnetic field are derived.

\section{Conclusions}
VHE emission from passive SMBH is plausible either via leptonic or hadronic processes. In order to detect this emission the giant elliptical galaxy NGC 1399 was observed by H.E.S.S. in 2005. NGC 1399 is not detected in these observations. The corresponding upper limit does not allow a firm estimation of the circumnuclear magnetic field. 


\subsubsection*{Acknowledgments}
The support of the Namibian authorities and of the University of Namibia in facilitating the construction and operation of H.E.S.S. is gratefully acknowledged, as is the support by the German Ministry for Education and Research (BMBF), the Max Planck Society, the French Ministry for Research, the CNRS-IN2P3 and the Astroparticle Interdisciplinary Programme of the CNRS, the U.K. Science and Technology Facilities Council (STFC), the IPNP of the Charles University, the Polish Ministry of Science and Higher Education, the South African Department of Science and Technology and National Research Foundation, and by the University of Namibia. We appreciate the excellent work of the technical support staff in Berlin, Durham, Hamburg, Heidelberg, Palaiseau, Paris, Saclay, and in Namibia in the construction and operation of the equipment. \\This work has been supported by the International Max Planck Research School (IMPRS) for Astronomy \& Cosmic Physics at the University of Heidelberg.

\bibliography{icrc0470}

\begin{thebibliography}{10}

\bibitem{gal-cen}
F.~{Aharonian} and A.~{Neronov}.
\newblock {High-Energy gamma rays from the massive black hole in the galactic
  center}.
\newblock {\em ApJ}, 619:306--313, 2005.

\bibitem{benb}
W.~{Benbow}.
\newblock {The H.E.S.S. Standard Analysis Technique}.
\newblock {\em Proceedings of Towards a Network of Atmospheric Cherenkov
  Detectors VII (Palaiseau)}, page 163, 2005.

\bibitem{berge}
{Berge}~D. et~al.
\newblock {Background modelling in very-high-energy $\gamma$-ray astronomy}.
\newblock {\em A$\&$A}, 466:1219, 2007.

\bibitem{lowen}
{Loewenstein}~M. et~al.
\newblock {Chandra Limits on X-Ray Emission Associated with the Supermassive
  Black Holes in Three Giant Elliptical Galaxies}.
\newblock {\em ApJ}, 555:L21--29, 2001.

\bibitem{mag}
{Magorrian}~A. et~al.
\newblock {The demography of massive dark objects in galaxy centers}.
\newblock {\em ApJ}, 115:2285--2305, 1998.

\bibitem{neron}
{Neronov}~A. et~al.
\newblock {TeV signatures of compact UHECR accelerators}.
\newblock {\em J. Exp. Theor. Phys.}, 100:656--662, 2004.

\bibitem{oconn}
{O' Connell} R.~W. et~al.
\newblock {UV/Optical Nuclear Activity in the gE Galaxy NGC 1399}.
\newblock {\em ApJ}, 635:305--310, 2005.

\bibitem{rich}
{Richstone}~D. et~al.
\newblock {Supermassive black holes and the evolution of galaxies.}
\newblock {\em Nature}, 395:A14, 1998.

\bibitem{sadl}
{Sadler} E.~M. et~al.
\newblock {Low-luminosity radio sources in early-type galaxies}.
\newblock {\em MNRAS}, 240:591, 1989.

\bibitem{fc}
G.~J. {Feldman} and R.~D. {Cousins}.
\newblock {Unified approach to the classical statistical analysis of small
  signals}.
\newblock {\em Phys. Rev. D}, 57:3873, 1998.

\bibitem{bick}
Z.~{Kuncic} and G.~V. {Bicknell}.
\newblock {Dynamics and Energetics of Turbulent, Magnetized Disk Accretion
  around Black Holes: A First-Principles Approach to Disk-Corona-Outflow
  Coupling}.
\newblock {\em ApJ}, 616:669, 2004.

\bibitem{lev}
A.~{Levinson}.
\newblock {Particle Acceleration and Curvature TeV Emission by Rotating,
  Supermassive Black Holes}.
\newblock {\em Phys. Rev. Let.}, 85:912--915, 2000.

\bibitem{pel}
S.~{Pellegrini}.
\newblock {Nuclear Accretion in Galaxies of the local Universe: Clues from
  CHANDRA Observations}.
\newblock {\em ApJ}, 624:155--161, 2005.

\bibitem{slane}
P.~{Slane} and S.~M. {Wagh}.
\newblock {TeV gamma-ray production in accreting black hole systems}.
\newblock {\em ApJ}, 364:198--202, 1990.

\end{thebibliography}
\bibliographystyle{plain}
\end{document}